\documentstyle[12pt]{article}
\begin{document}
\title{A Derivation of the Correct Treatment of ${1\over{(\eta\cdot k)^p}}$
Singularities in Axial Gauges}
\author{{Satish. D. Joglekar}\thanks{e-mail:sdj@iitk.ac.in},
{A. Misra} \thanks{e-mail:aalok@iitk.ac.in}\\
Department of Physics, Indian Institute of Technology,\\
 Kanpur 208 016, UP, India}
\maketitle
\begin{abstract}
We use the earlier results on the correlations of axial gauge Green's
functions and the Lorentz gauge Green's functions 
obtained via finite field-dependent BRS transformations to 
study the question of the correct treatment of ${1\over{(\eta\cdot k)^p}}$-
type singularities in the axial gauge boson propagator.
We show how the known  treatment 
of  the ${1\over{(k^2)^n}}$-type singularity in
the Lorentz-type gauges can be used to write down the axial
propagator via field transformation. We examine 
the singularity structure of the
latter and find that the axial propagator so constructed has
$no$ spurious poles, but a complex structure near $\eta\cdot k=0$.
\end{abstract}

The known high energy physics is well represented
by the Standard Model (SM): a nonabelian gauge theory.
Hence, the importance of Feynman diagrams calculations
in nonabelian gauge theories need not be elaborated. Among the gauges commonly
used in such calculations 
are the Lorentz-type and the axial-type gauges.
The former are commonly used on account of simplicity of Feynman rules,
Lorentz covariance and possibility of testing
gauge-independence. Axial gauges have also been frequently used in
SM calculations on account of  the ghost-free nature
which  cuts down the number of Feynman diagrams to be evaluated.  
Axial gauges suffer from lack of manifest covariance and more
importantly  the spurious
${1\over{(\eta\cdot k)^p}}$ singularities in  the
propagators. Various
treatments for such singularities have been
proposed such  as Principle Value Prescription
(PVP)\cite{pvp} and the Mandelstam-Leibbrandt (ML) prescription \cite{lm}.
They are of an ad-hoc nature and while they are successful
in many calculations, they 
lead to a number of difficulties \cite{diff} especially for the
light-cone gauge (LCG).
A prescription for  gauges of  the form $A_1+\lambda A_3=0$
(LCG not included) has also been derived using canonical
quantization \cite{landsh}.

We approach the problem of interpreting
the ${1\over{(\eta\cdot k)^p}}$-type singularities in the
{\it path-integral} formulation in a different way. Unlike the PVP
and the LM approach, our approach is to $derive$ the treatment of axial gauge
propagator poles (and in  fact attempt to obtain the correct way to do
axial gauge calculations) by using the connection
of axial gauge Green's functions
with the corresponding Lorentz gauge calculation
established earlier in \cite{BRS1}. 
Unlike \cite{landsh}, our approach includes LCG also.
We have number of motivations for doing so.
We utilize the approach in \cite{BRS1} which establishes
explicitly connections between
Green's functions in Lorentz-type and axial-type gauges.
Among theoretical motivations is to establish
identity of physical observables in a totally distinct set of gauges. On
practical side, we would like to remove discrepancies reported in
observable anomalous dimension calculations [see references in \cite{BRS1}].
We would also like to place axial gauge calculations on a rigorous theoretical
foundation and remove any problems associated with various 
prescriptions \cite{BRS1}.  
The present work provides a step in these directions.

The basic idea (as also outlined in \cite{jb,talk}) is as follows.
In Lorentz-type gauges, there are also spurious singularities at $k^2=0$
(except the Feynman gauge). We correctly deal with these 
by the $k^2\rightarrow k^2+i\epsilon$
treatment, which amounts to an addition of a terms
 $-i\epsilon(A^2/2-{\bar c}c)$ to the action.
As is well known, the $-i\epsilon A^2/2$ term provides a damping in the 
Minkowskian formulation of the path integral
for the generating functional of Lorentz-type gauge theories
for the transverse modes.  We now propose to use this
well-defined treatment in Lorentz-type gauges by performing 
a field transformation (on gauge and ghost fields)
that  converts the Lorentz and axial gauge generating
functionals. Such a field transformation has been established
in \cite{jb} based on an earlier work \cite{jm} and is a
field-dependent generalization of  BRS transformation called Finite
Field-dependent BRS (FFBRS) transformation.
(It is also reproduced below in (\ref{eq:result1})  
The idea of such a transformation
was used to correlate arbitrary Green's functions in
axial-type gauges to  those
in Lorentz-type gauges in \cite{BRS1}. The result 
reads:
\begin{eqnarray}
\label{eq:result3}
& & \langle O[\phi]
\rangle_A\equiv\int{\cal D}\phi^\prime O[\phi^\prime]
e^{iS_{\rm eff}^A[\phi^\prime]}\nonumber\\
& & =\int{\cal D}\phi{\cal O}[\phi]e^{iS^L_{\rm eff}[\phi]}+
\int{\cal D}\phi\sum_i\delta\phi_i[\phi]
{\delta{\cal O}\over{\delta\phi_i}} e^{iS^L_{\rm eff}},
\end{eqnarray}
where the summation over $i$ runs over fields $A, c, {\bar  c}$ and
\begin{eqnarray}
\label{eq:result1}
& & \phi^\prime=\phi+\biggl(\tilde\delta_1[\phi]\Theta_1[\phi]
+\tilde\delta_2[\phi]\Theta_2[\phi]\biggr)\Theta^\prime[\phi]
\nonumber\\
& & \equiv\phi+\delta\phi[\phi]
\end{eqnarray}
is an FFBRS \cite{jm} with
\begin{equation}
\label{eq:Theta12def}
\Theta_{1,2}[\phi]\equiv\int_0^1 d\kappa (1,\kappa)exp\biggl(\kappa f_1[\phi]
+{\kappa^2\over 2}f_2[\phi]\biggr);
\end{equation}
\begin{equation}
\label{eq:f12def1}
f[\tilde\phi,\kappa]\equiv 
f_1[\tilde\phi]+\kappa f_2[\tilde\phi];
\end{equation}
\begin{eqnarray} 
\label{eq:f12def2}
& & f_1[\phi]\equiv i\int d^4x\biggl[{\partial\cdot A^\alpha\over\lambda}
(\partial\cdot A^\alpha-\eta\cdot A^\alpha)+{\bar c}
(\partial\cdot{\rm D}-\eta\cdot{\rm D})c\biggr]
\nonumber\\
& & f_2[\phi]\equiv -{i\over\lambda}
\int d^4x (\partial\cdot A^\alpha-\eta\cdot A^\alpha)^2,
\end{eqnarray}
and
\begin{equation}
\label{Thprimedef}
\Theta^\prime\equiv i\int d^4x {\bar c}^\alpha
(\partial\cdot A^\alpha-\eta\cdot A^\alpha).
\end{equation}
An alternate and more effective expression can be given \cite{BRS1}:
\begin{equation}
\label{eq:AtoL2}
\langle {\cal O}\rangle_A=\langle{\cal O}\rangle_L+\int_0^1 d\kappa\int D\phi  
\sum_i\biggl(\tilde\delta_{1,i}[\phi]+\kappa\tilde\delta_{2,i}[\phi]
\biggr)\Theta^\prime[\phi]
{\delta{\cal O}\over{\delta\phi_i}} e^{iS^M_{\rm eff}},
\end{equation}
where $\tilde\delta_{1,i}$ and $\tilde\delta_{2,i}$ are defined in \cite{BRS1}
via
\begin{equation}
\label{eq:tildedelsdef}
\tilde\delta_{\rm BRS}\phi_i\equiv
\biggl(\tilde\delta_{1,i}+\kappa\tilde\delta_{2,i}\biggr)\delta\Lambda
\end{equation}
where $\tilde\delta_{\rm BRS}\phi_i$ are the BRS variations for the mixed
gauge function
[$\partial\cdot A(1-\kappa)+\kappa\eta\cdot A$].
The basic idea is to use (\ref{eq:AtoL2}) to relate the axial and
Lorentz gauge propagators. The only shortcoming of the above relation
is that it does not include the
$i[-\epsilon A^2/2+\epsilon {\bar c}c]$ terms in the Lorentz gauge effective action.
The modification  of (\ref{eq:AtoL2}) due to this term  can be obtained
using the reverse of (\ref{eq:result3}) itself. We first do this below.

The Lorentz gauge vacuum-to-vacuum amplitude $W$, with
$\epsilon$ terms included reads:
\begin{equation}
\label{eq:eps1}
W^L=\int{\cal D}\phi e^{iS_{\rm eff}^L[\phi]+i\epsilon O_1[\phi]}
\end{equation}
with $O_1\equiv i\int d^4x \biggl(-A^2/2+{\bar c}c\biggr)$. We look
upon $e^{i\epsilon O_1}$ as $O$ in (\ref{eq:result3}) to arrive at:
\begin{eqnarray}
\label{eq:esp2}
& & W^L=\int {\cal D}\phi^\prime
e^{iS_{\rm eff}^A}\biggl(O[\phi^\prime]
+\sum_i\delta\phi_i[\phi^\prime]{\delta O\over{\delta\phi_i}}\biggr)
\nonumber\\
& & \equiv
\int {\cal D}\phi^\prime e^{iS_{\rm eff}^A+i\epsilon O^\prime_1}\equiv
W^A
\end{eqnarray}
where
\begin{equation}
\label{eq:eps3}
O_1^\prime[\phi^\prime]=O_1[\phi]+\delta\phi_i[\phi]
{\delta O_1\over{\delta\phi_i}}[\phi],
\end{equation}
and gives the $correct$ $\epsilon$ terms to be added to
$S_{\rm eff}^A[\phi^\prime]$. The axial gauge Green's functions are then
computed with the $\epsilon$-dependent effective action
$S_{\rm eff}^A[\phi^\prime]+\epsilon O_1^\prime[\phi^\prime]$, viz:
\begin{equation}
\label{eq:eps4}
\langle O[\phi^\prime]\rangle_A=\int{\cal D}\phi^\prime O[\phi^\prime]
e^{iS_{\rm eff}^A[\phi^\prime]+i\epsilon O^\prime_1[\phi^\prime]}.
\end{equation}
It turns out to be unnecessary to know the form of
$O_1^\prime$ explicitly \cite{inprep}. One can relate the Green's function
$\langle O[\phi^\prime]\rangle$ to
the corresponding Lorentz gauge quantities by the procedure
established in \cite{BRS1} and the result is simply:
\begin{eqnarray}
\label{eq:presp15}
& & \langle O\rangle_A = \langle O\rangle_L\nonumber\\
& & +\int_0^1 d\kappa\int{\cal D}\phi e^{iS^M_{\rm eff}[\phi,\kappa]
-i\epsilon(A^2/2-{\bar c}c)}\biggl[\biggl(\delta_1[\phi]
+\kappa\delta_2[\phi]\biggr)\Theta^\prime[\phi]
{\delta O\over{\delta\phi}}\biggr].
\end{eqnarray}

Thus, in {\it this form}, the only effect on the second term is 
to modify $S_{\rm eff}^M$ by
$-\epsilon\int (A^2/2-{\bar c}c)d^4x$ inside $\kappa$-integration.

We now employ the result (\ref{eq:presp15}) for the propagators. We 
set $O[\phi]= A^\alpha_\mu(x)A^\beta_\nu(y)$. The equation
(\ref{eq:presp15}) then reads:
\begin{eqnarray}
\label{eq:exres}
& & iG^{0A\ \alpha\beta}_{\mu\nu}(x-y)=iG^{0L\ \alpha\beta}_{\mu\nu}(x-y)
+i\int_0^1 d\kappa\int{\cal D}\phi e^{iS_{\rm eff}^M[\phi,\kappa]-i
\epsilon\int(A^2/2-{\bar c}c)d^4x}\nonumber\\
& & \times
\biggl(({\rm D}_\mu c)^\alpha(x)A^\beta_\nu(y)
+A^\alpha_\mu(x)
({\rm D}_\nu c)^\beta(y)\biggr)
\int d^4z{\bar c}^\gamma(z)(\partial\cdot A^\gamma-\eta\cdot A^\gamma)(z).
\nonumber
\end{eqnarray}
This leads to, for zero loop case,
\begin{eqnarray}
\label{eq:exres1}
&  & G^{0 A\ \alpha\beta}_{\mu\nu}(x-y)=G^{0 L\ \alpha\beta}_{\mu\nu}(x-y)
\nonumber\\
& & -i\int_0^1 d\kappa\biggl[-i\partial_\mu^x
\tilde G^{0M}(x-y)(\partial_z^\sigma-\eta^\sigma)
\tilde G^{0M\ \alpha\beta}_{\sigma\nu}
+(\mu,x,\alpha)\leftrightarrow(\nu,y,\beta)\biggr].
\nonumber\\
& & 
\end{eqnarray}
The last term on  the right hand side  involves $\kappa$-dependent functions for ghost and 
gauge fields:
\begin{equation}
\label{eq:eps5}
\tilde G^{0 M}(x-y)
=\int d^4q{e^{-iq\cdot (x-y)}\over{(\kappa-1)q^2-i\kappa q\cdot\eta-i\epsilon}}
\end{equation}
and
\begin{equation}
\label{eq:eps6}
\tilde G^{0 M\ \alpha\beta}_{\sigma\nu}(x-y)
=\delta^{\alpha\beta}\int d^4k e^{-ik\cdot (x-y)} \tilde G^{0 M}_{\sigma\nu}(k)
\end{equation}
with
\begin{eqnarray}
\label{eq:pres6}
& & \tilde G^{0 M}_{\mu\rho}(k)
=-{1\over{k^2+i\epsilon}}\Biggl[g_{\mu\rho}+\nonumber\\
& & {{\biggl(\biggl[[(1-\kappa)^2-\lambda]-{\eta^2\kappa^2\over{k^2+i\epsilon}}
\biggr]k_\mu k_\rho
-i\kappa(1-\kappa)k_{[\mu}\eta_{\rho]}
+{\kappa^2\eta\cdot k\over{k^2+i\epsilon}}k_{[\mu}\eta_{\rho]_+}
+i{\kappa^2\epsilon\over{k^2+i\epsilon}}\eta_\mu\eta_\rho\biggr)}
\over{\biggl(-{\kappa^2\over{(k^2+i\epsilon)}}
\biggl[(\eta\cdot k)^2-\eta^2k^2+(k^2+\eta^2)(k^2+i\epsilon)\biggr]
+2k^2\kappa
-i\epsilon\lambda-k^2\biggr)}}\Biggr].
\nonumber\\
& & 
\end{eqnarray}
[It should be emphasized that (\ref{eq:eps5}), (\ref{eq:eps6}) are only
intermediate objects occurring in calculations
and are $not$ the actual   ghost   and gauge
propagators (even in intermediate gauges) 
as the latter must be evaluated ultimately with
a term like $\epsilon O^\prime_1[\phi^\prime,\kappa]$ in the exponent.] We
obtain:
\begin{eqnarray}
\label{eq:Result1}
& & \tilde G^{0 A}_{\mu\nu}-\tilde G^{0 L}_{\mu\nu}
={-i\over{(k^2+i\epsilon)^2(1-i\xi_1-i\xi_2)(1-i\xi_2+\xi_1^2+i\xi_2\xi_3)}}
\nonumber\\
& & \times\int_0^1 d\kappa{\Biggl[k_\mu k_\nu\biggl(\kappa
+\biggl[{i\lambda-\xi_1(1-\lambda)\over{\xi_1+i\xi_3}}\biggr]\biggr)
(\xi_1+i\xi_3)+\eta_\mu k_\nu
\biggl(\kappa+\biggl[{1-i\xi_2(1-\lambda)\over{-1-i\xi_1+i\xi_2}}\biggr]\biggr)
(-1-i\xi_1+i\xi_2)\Biggr]
\over{(\kappa-a_1)(\kappa^2-2\gamma\kappa+\beta)}}\nonumber\\
& & +(k\rightarrow-k,\ \mu\leftrightarrow\nu) 
\end{eqnarray}
with
\begin{eqnarray}
\label{eq:defins}
& & \xi_1\equiv{\eta\cdot k\over{k^2+i\epsilon}};\nonumber\\
& & \xi_2\equiv{\epsilon\over{k^2+i\epsilon}};\nonumber\\
& & \xi_3\equiv{\eta^2\over{k^2+i\epsilon}};\nonumber\\
& & a_1\equiv{1\over{1-i\xi_1-i\xi_2}};\nonumber\\
& & \gamma\equiv{(1-i\xi_2)\over{1-i\xi_2+\xi_1^2+i\xi_2\xi_3}}\equiv{1-i\xi_2\over D};\nonumber\\
& & \beta\equiv{1+i\xi_2(\lambda-1)\over{1-i\xi_2+\xi_1^2+i\xi_2\xi_3}}
=\gamma +{i\xi_2\lambda\over D}.
\end{eqnarray}
The quadratic in the denominator 
can be rewritten as $(\kappa-\kappa_1)(\kappa-\kappa_2)$ with
\begin{eqnarray}
\label{eq:kappa12def}
& & \kappa_{1,2}=\gamma\pm\sqrt{\gamma^2-\beta}
={{1-i\xi_2}\pm
\sqrt{(1-i\xi_2)^2
-[1+i\xi_2(\lambda-1)](1-i\xi_2+\xi_1^2+i\xi_2\xi_3)}\over D}\nonumber\\
& & \equiv{1-i\xi_2\pm\sqrt{Y}\over D}.
\end{eqnarray}
We  note that of  the three zeros of the denominators, 
two are equal are $\epsilon=0$, since
\begin{eqnarray}
\label{eq:a1kappa1}
& & \kappa_1|_{\epsilon=0}
={1\over{1+\xi_1^2}}+\sqrt{\biggl(
{1\over{1+\xi_1^2}}\biggr)^2-{1\over{1+\xi_1^2}}}\nonumber\\
& & ={1+i\xi_1\over{1+\xi_1^2}}=a_1|_{\epsilon=0}.
\end{eqnarray}
We shall now state an important convention in defining the square roots in
(\ref{eq:kappa12def}). The square root $\sqrt{Y}$ has branch points 
at $\pm\sqrt{{-i\xi_2[(1-i\xi_2)\xi_3+\lambda(1-i\xi_2+i\xi_2\xi_3)]
\over{1-i\xi_2(1-\lambda)}}}$ and these lie a distance O$(\sqrt{\epsilon})$
away from the origin [For LCG, in the $k^2=0$ subspace, 
however, $\sqrt{Y}=i\sqrt{\lambda}\xi_1$
has no branch cut in $\xi_1$-plane]. We choose
the branch cut joining these. To obtain the value of 
$+\sqrt{Y}$ at any point $\xi^\prime$ 
not on the branch cut, we consider $\sqrt{Y}$
for $\xi_1=M\xi^\prime$ as $M\to+\infty$. 
Then we can ignore $\epsilon$ terms in this
case and $\sqrt{Y}=\sqrt{-\xi_1^2}(k^2\neq0$)
or $\sqrt{-\lambda\xi_1^2}(k^2=0$). These
we define to be $i\xi_1$ or $i\sqrt{\lambda}\xi_1$ respectively
($\lambda>0$ assumed).  
We then define $\sqrt{Y}$ for $\xi_1=\xi^\prime$ by requiring that
the phase of $\sqrt{Y}$ is a continuous function of $M$ for $1\leq M<\infty$.
From this and from the fact that $Y\equiv Y(\xi_1^2)$, we learn that
$\sqrt{Y}(-\xi_1)=-\sqrt{Y}(\xi_1)$. Hence, $\kappa_2(-\xi_1)=\kappa_1(\xi_1)$.
We further note that this prescription defines
uniquely $\sqrt{Y}$ for  real $\eta\cdot k\neq 0$ since  the branch cut  cuts
the real $\eta\cdot k$ axis only at the
origin $\eta\cdot k=0$.

We further note that both the $k_\mu k_\nu$
and the $\eta_\mu k_\nu$ terms involve an integral of the same form
\begin{equation}
\label{eq:kappaint}
\int_0^1 d\kappa 
{(\kappa+\alpha)\over{(\kappa-a_1)(\kappa-\kappa_1)(\kappa-\kappa_2)}},
\end{equation}
the constant $\alpha$ being different for the $k_\mu k_\nu$ and 
$\eta_\mu k_\nu$ terms.
This can be evaluated and reorganized as:
\begin{eqnarray}
\label{eq:pres10}
& & {(a_1+\alpha)\over{(a_1-\kappa_1)(a_1-\kappa_2)}}
ln\biggl[{{1-a_1}\over{-a_1}}\biggr]\nonumber\\
& & +{(\kappa_1+\alpha)\over{(\kappa_1-a_1)(\kappa_1-\kappa_2)}}
ln\biggl[{{1-\kappa_1}\over{-\kappa_1}}\biggr]
+{(\kappa_2+\alpha)\over{(\kappa_2-a_1)
(\kappa_2-\kappa_1)}}ln\biggl[{{1-\kappa_2}\over{-\kappa_2}}\biggr]
\nonumber\\
& & \equiv {1\over{a_1-\kappa_2}}\biggl({(\alpha+a_1)\over{(a_1-\kappa_1)}}
ln\biggl[{{\kappa_1-\kappa_1a_1}\over{a_1-a_1\kappa_1}}\biggr]
-{(\kappa_2+\alpha)\over{(\kappa_1-\kappa_2)}}
ln\biggl[{{\kappa_2-\kappa_1\kappa_2}
\over{\kappa_1-\kappa_1\kappa_2}}\biggr]\biggr).
\end{eqnarray}
In the second term, we note that under $\xi_1\to-\xi_1$,
\begin{equation}
\label{eq:interch1}
{1\over{\kappa_1-\kappa_2}}ln\biggl({\kappa_2-\kappa_2\kappa_1
\over{\kappa_1-\kappa_2\kappa_1}}\biggr)
\end{equation}
is invariant under as $\kappa_1\leftrightarrow\kappa_2$. Using
this and the necessary Bose symmetrization implicit in
(\ref{eq:Result1}),
it can be shown  that \cite{inprep} this term does not contribute to 
(\ref{eq:Result1}) as $\epsilon\to 0$. [In the LCG, however,
special care needs to be exercised in the subspace $\eta\cdot k=k^2=0$.
We must follow one of two options: (i) take limit $\lambda\to 0$ in the
end of calculation, or (ii) interpret LCG as the limit $\eta^2\to 0$ 
taken in the very end of the calculation.
(Then we may set $\lambda=0$ in the beginning)]
Hence the propagator (\ref{eq:Result1}) is given in terms of the first term
in (\ref{eq:pres10}):
\begin{equation}
\label{eq:mainln}
{a_1+\alpha\over{(a_1-\kappa_2)(a_1-\kappa_1)}}
ln\biggl[{{\kappa_1-\kappa_1a_1}\over{a_1-a_1\kappa_1}}\biggr]
\end{equation}
substituted for (\ref{eq:kappaint}) in (\ref{eq:Result1}). Hence, we shall study the
structure of (\ref{eq:mainln}) in detail.
The singularity structure of (\ref{eq:mainln}) is dependent on the 
denominators and the logarithm. The equation
(\ref{eq:mainln}), in general reads:
\begin{equation}
\label{eq:lnP}
{(a_1+\alpha)D(1-i\xi_1-i\xi_2)^2
\over{i\xi_2(1-\lambda)P(\xi_1)}}ln\biggl[{-i(\xi_1+\xi_2)
\over{{-i\xi_2\lambda\over{1-i\xi_2(1-\lambda)}}
-\sqrt{[{1-i\xi_2\over{1-i\xi_2(1-\lambda)}}]^2-{D\over{1-i\xi_2(1-\lambda)}}}}}
\biggr]
\end{equation}
with
\begin{equation}
\label{eq:defP}
P(\xi_1)\equiv
\xi_1^2+2i\xi_1(1-i\xi_2)+{\lambda+i\xi_2(1-2\lambda)
+\xi_2^2(1-\lambda)+\xi_3\over{1-\lambda}}.
\end{equation}
The apparent complexity of (\ref{eq:lnP})
actually exists only in the small region of the $\eta\cdot k$ complex
plane near the  origin. We note  that for 
$|a_1-\kappa_1|<|a_1(1-\kappa_1)|$, the 
expression
(\ref{eq:mainln}) can be expressed as 
\begin{equation}
\label{eq:expln}
{1\over{a_1-\kappa_1}}
ln\biggl[{\kappa_1-a_1\kappa_1\over{a_1-a_1\kappa_1}}\biggr]
=-{1\over{a_1-a_1\kappa_1}}
+O(a_1-\kappa_1).
\end{equation}
The condition $|a_1-\kappa_1|<|a_1(1-\kappa_1)|$ implies
\begin{equation}
\label{eq:condition}
Im\biggl({\sqrt{-(\eta\cdot k)^2-i\epsilon\eta^2}\over
{\sqrt{{k^2\over{k^2+i\epsilon}}}(\eta\cdot k+\epsilon)}}\biggr)>{1\over 2}
\end{equation}
and this covers all of real $\eta\cdot k$ axis save the region
$(-\epsilon,0)$ for $\eta^2\neq0$ and $(-\epsilon,\epsilon)$ for LCG. Thus,
neglecting O$(\epsilon)$ terms
(\ref{eq:expln}) reads:
\begin{equation}
\label{eq:expln2}
-{1\over{a_1(1-\kappa_1)(a_1-\kappa_2)}}.
\end{equation}
For $|\eta\cdot k|>>\epsilon$, this is easily seen to be
\begin{equation}
\label{eq:smalln.k}
-{(1-i\xi_1)^2(1+\xi_1^2)\over{2\xi_1^2}}
\end{equation}
and leads
 to the usual behavior of the axial propagator
when substituted into
(\ref{eq:Result1}), which then reads:
\begin{equation}
\label{eq:eps=0}
\tilde G^{0 A}_{\mu\nu}-\tilde G^{0 L}_{\mu\nu}= 
-{1\over k^2}k_\mu k_\nu
\biggl({(\lambda k^2+\eta^2)\over{(\eta\cdot k)^2}}+{(1-\lambda)\over k^2}
\biggr) + {k_{[\mu}\eta_{\nu]_+}\over{k^2\eta\cdot k}}.
\end{equation}
We now turn to the analytic structure of (\ref{eq:lnP}) near
$\xi_1=0$ on the real $\eta\cdot k$ axis. We note:
(i)  $P(\xi_1)$ has no zeros on the real $\eta\cdot k$ axis.
[For the LCG, we need to exercise care in the $k^2=0$ subspace.
Here we need to take the limit $\eta^2\to 0$ at the end to avoid
singularities in this subspace.] (ii) The apparent complexity of 
(\ref{eq:lnP}) is substantially
 reduced if we set $\lambda=0$
[which is the limit we mean to take anyhow]. Then (\ref{eq:lnP}) reads:
\begin{eqnarray}
\label{eq:lnP2}
& & {a_1+\alpha\over{i\xi_2[\xi_1^2+2i\xi_1(1-i\xi_2)+i\xi_2(1-i\xi_2)+\xi_3]}}
ln\biggl[{i(\xi_1+\xi_2)\sqrt{1-i\xi_2}\over{\sqrt{-\xi_1^2-i\xi_2\xi_3}}}\biggr]
\nonumber\\
& & ={(a_1+\alpha)(k^2+i\epsilon)^3\over{i\epsilon[(\eta\cdot k)^2+2ik^2\eta\cdot k
+i\epsilon k^2+\eta^2(k^2+i\epsilon)]}}
ln\biggl[{i(\eta\cdot k+\epsilon)\over{\sqrt{-(\eta\cdot k)^2-i\epsilon\eta^2}}}
\sqrt{{k^2\over{k^2+i\epsilon}}}\biggr].\nonumber\\
& & 
\end{eqnarray}
The above expression can be used for $\eta^2\neq0$. For LCG, we need to exercise 
care in the
$k^2=0$ subspace. We may either  (i) take the limit $\eta^2\to 0$ in the
end or (ii) we may keep $\lambda$ small
in $P(\xi_1)$ and express (\ref{eq:lnP}) as
\begin{eqnarray}
\label{eq:lnP3}
& & {a_1+\alpha\over{i\xi_2[\xi_1^2+2i\xi_1(1-i\xi_2)+i\xi_2(1-i\xi_2)+\lambda]}}
ln\biggl[{i(\eta\cdot k+\epsilon)\over{i(\eta\cdot k)}}
\sqrt{{k^2\over{k^2+i\epsilon}}}\biggr]
\nonumber\\
& & ={(a_1+\alpha)(k^2+i\epsilon)^3\over{i\epsilon[(\eta\cdot k)^2+2ik^2\eta\cdot k
+i\epsilon k^2+\lambda(k^2+i\epsilon)^2]}}
ln\biggl[{i(\eta\cdot k+\epsilon)\over{i(\eta\cdot k)}}
\sqrt{{k^2\over{k^2+i\epsilon}}}\biggr].\nonumber\\
& & 
\end{eqnarray}
Expressions (\ref{eq:lnP2}) and (\ref{eq:lnP3}) have
a (mild) logarithmic singularity at $\eta\cdot k=-\epsilon$ and the
expression (\ref{eq:lnP3}) has in addition a logarithmic singularity
at $\eta\cdot k=0$. Thus, the singularity structure of the propagator is
softened.

We summarize the view presented in this work 
briefly:

(i) The propagator in axial gauge, naively calculated, has spurious
singularities.

(ii) The correct treatment of these singularities
is obtained by relating this propagator to the
corresponding Lorentz gauge treatment. This can be done
by using the FFBRS.

(iii) The propagator of (\ref{eq:Result1}) gives, however complex, the
actual correct treatment of these singularities.

(iv) While for $|\eta\cdot k|>>\epsilon$, it
gives the usual propagator, the actual analytic nature of the propagator,
in the vicinity of the origin is much more 
complicated than indicated by various
prescriptions suggested earlier.

(v) We may expect that the ills associated with the axial/LC gauge
\cite{diff,landsh} may be cured if the structure presented here is  taken into 
account. This will be left to a later work.

We finally summarize our results. We find:
\begin{equation}
\label{eq:freslt1}
\tilde G^{0 A}_{\mu\nu}=\tilde G^{0 L}_{\mu\nu}\
+\biggl[\biggl(k_\mu k_\nu\Sigma_1+\eta_\mu k_\nu\Sigma_2\biggr)ln\Sigma_3
+(k\to -k;\mu\leftrightarrow\nu)\biggr]
\end{equation}
where
\begin{eqnarray}
\label{eq:freslt2}
& & \Sigma_1\equiv {-(k^2-i\eta\cdot k)\biggl({\eta\cdot k+i\eta^2
\over{k^2-i\eta\cdot k}}+i\lambda
-{(1-\lambda)\eta\cdot k\over{k^2+i\epsilon}}\biggr)
\over{\epsilon\Sigma}}\nonumber\\
& & \Sigma_2\equiv {-(k^2-i\eta\cdot k)\biggl(
-\biggl[{k^2+i\eta\cdot k\over{k^2-i\eta\cdot k}}
\biggr]+1-{i\epsilon(1-\lambda)\over{k^2+i\epsilon}}\biggr)
\over{\epsilon\Sigma}}\nonumber\\
\nonumber\\
& & \Sigma_3\equiv{-i(\eta\cdot k+\epsilon)(k^2+i\epsilon\lambda)
\over{(k^2+i\epsilon)\biggl(-i\epsilon\lambda-\sqrt{k^4-(k^2+i\epsilon\lambda)
\biggl[k^2+{(\eta\cdot k)^2+i\epsilon\eta^2\over{k^2+i\epsilon}}\biggr]}\biggr)}},\nonumber\\
& & {\rm and}\nonumber\\
& & \Sigma\equiv\biggl[(1-\lambda)[(\eta\cdot k)^2+2ik^2\eta\cdot k]
+i\epsilon k^2(1-2\lambda)+\lambda(k^2+i\epsilon)^2+\eta^2(k^2+i\epsilon)\biggr].
\nonumber\\
& & 
\end{eqnarray} 
We leave the elaborate discussion to a detailed publication \cite{inprep}.

It is possible to show that in the region around the point $\eta\cdot k=0$,
the propagator can be replaced {\it effectively} by a much simpler
expression.
(We shall leave the details to
Reference [9]). We show that the $k^0$ integration over this
propagator can be replaced by a $k^0$-integration over (most of) the real
axis combined over semicircle in the LHP of radius $>>\sqrt{\epsilon}$
(where the complication due to presence of $\epsilon $ can
be dropped and the usual simple form can be used)
and an  additional effective term of much simpler form that rounds up
effectively the  complex structure near $\eta\cdot k=0$.
For $\eta^2\neq0$, and $k^2\neq0$, the latter reads
\begin{equation}
\label{eq:fresult3}
\delta\biggl(k^0-{1\over2}\sqrt{{\epsilon\eta^2\over i}}
-\vec\eta\cdot\vec k\biggr)
\biggl[k_\mu k_\nu D_1+\eta_\mu k_\nu D_2+\eta_\nu k_\mu D_3\biggr]
+\mu\leftrightarrow\nu;k\rightarrow-k
\end{equation}
\begin{eqnarray}
\label{eq:fresult4}
& & D_1\equiv -{\pi\eta^2\over{\cal K}_1}
i\sqrt{{i\eta^2\over\epsilon}}
+{i\pi(\eta^2)^2{\cal K}_2\over{2{\cal K}_1^2}}
;\nonumber\\
& & D_2\equiv {i\pi\eta^2\over{2{\cal K}_1}};\nonumber\\
& & D_3\equiv
-{i\pi\eta^2\over{2{\cal K}_1}},
\end{eqnarray}
where
\begin{eqnarray}
\label{eq:simpeq14}
& & {\cal K}_1\equiv\biggl((\vec\eta\cdot\vec k)^2-\vec k^2\biggr)
(\eta^2+i\epsilon);
\nonumber\\
& & {\cal K}_2\equiv 2i\biggl((\vec\eta\cdot\vec k)^2-\vec k^2\biggr)
+2\vec\eta\cdot\vec k(\eta^2+i\epsilon).
\end{eqnarray}
We further note {\it that if we define the LCG as the $\eta^2\rightarrow0$
limit, then this additional term (\ref{eq:fresult3}) vanishes}. Thus,
we obtain a simple result of the LCG. For details, 
refer to \cite{inprep}.

\end{document}